\documentclass[11pt]{article}
\usepackage{moriond,epsfig, subfigure}

\def\be{\begin{equation}}
\def\ee{\end{equation}}
\def\bea{\begin{eqnarray}}
\def\eea{\end{eqnarray}}

\begin{document}
\vspace*{4cm}
\title{Measurements of the polarisation amplitudes and triple product asymmetries in {$B_s^0 \to \phi \phi$}}

\author{ D. LAMBERT, on behalf of the LHCb Collaboration }

\address{School of Physics and Astronomy, James Clerk Maxwell Building, Mayfield Road, \\ Edinburgh EH9 3JZ, Scotland
\begin{scriptsize}
•
\end{scriptsize}}
\maketitle\abstracts{
Using 1~fb$^{-1}$ of $pp$ collision data collected at center of mass energy $\sqrt{s} = 7$ TeV during 2011 by the LHCb detector. Measurements of the triple product asymmetries, polarisation amplitudes and strong phase difference in the decay {$B_s^0 \to \phi \phi$} are presented.
}
\section{Introduction}
In the Standard Model the decay {$B_s^0 \to \phi \phi$} proceeds via a flavour changing neutral current process. In such processes contributions from beyond the Standard Model are possible via the introduction of new amplitudes or phases in the penguin loop. Hence studies of the polarisation amplitudes and triple product asymmetries in this mode provide important tests of the Standard Model.~\cite{gr1,ben1,dat1}

The decay {$B_s^0 \to \phi \phi$} is a pseudoscalar to vector-vector transition. Therefore, there are three possible spin configurations of the vector mesons allowed by conservation of angular momentum. These can be written as linear polarisation states $A_0$, $A_\parallel$ and $A_\perp$.
The final state is a superposition of $CP$-even and $CP$-odd states. The longitudinal($A_0$) and parallel($A_\parallel$) components are $CP$-even while the perpendicular($A_\perp$) component is $CP$-odd.

Tree dominated decays such as $B^0 \to \rho^+ \rho^-$ are prominently longitudinally polarised~\cite{Aubert:2003xc} while in penguin dominated decays such as $B^0 \to \phi K^{*0}(892)$ and $B^0 \to \rho^0 K^{*0}(892)$, roughly equal longitudinal and transverse components are observed.~\cite{aubert2,aubert1,chen1} The predictions 
do suffer from large hadronic uncertainties.~\cite{benk1,cheng1,ali1}
\begin{figure}[b]
\vspace{-10mm}
\setlength{\unitlength}{1mm}
\begin{center}
  \begin{picture}(90,45)
    \put(0,-1){
      \includegraphics*[width=90mm]{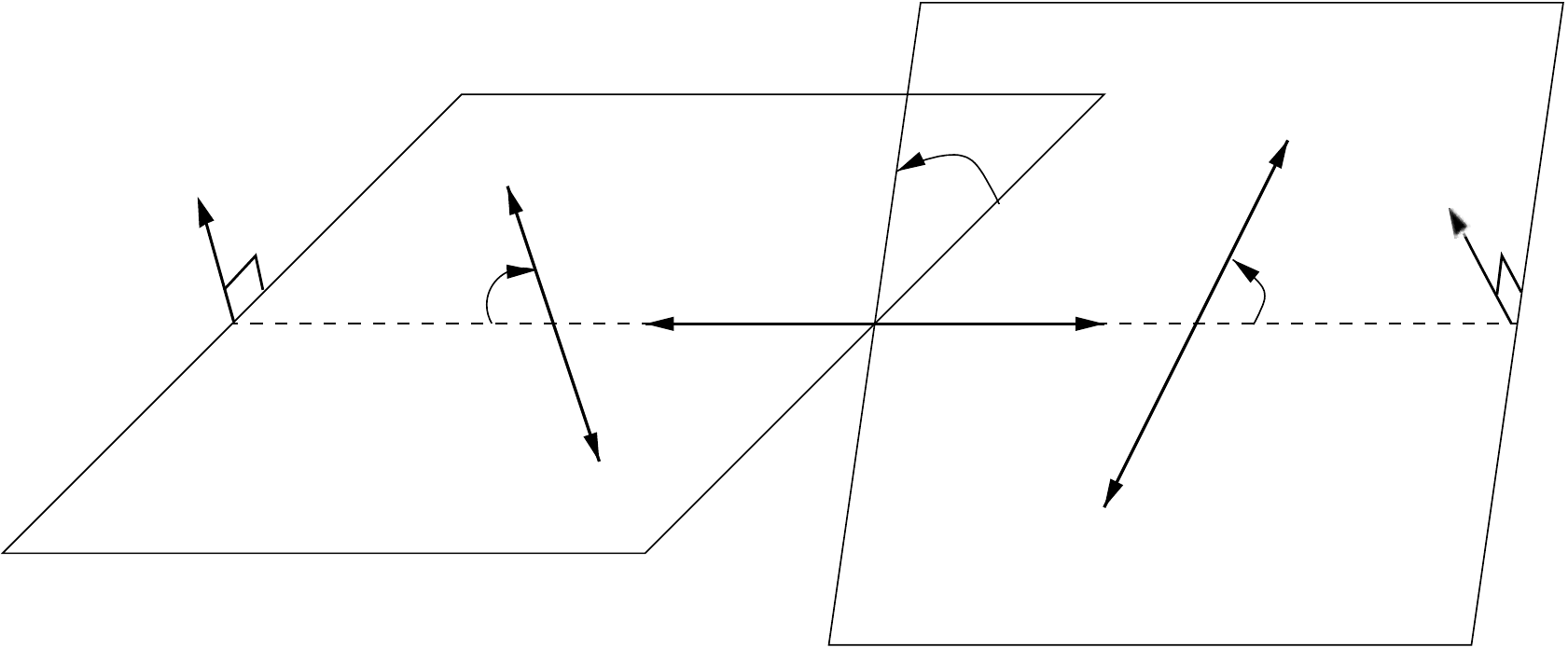}
    }
    \scriptsize
      \put(52,15){$B^0_s$}
      \put(54,23){$\Phi$}
      \put(24,19){$\theta_{1}$}
      \put(75,19){$\theta_{2}$}
      \put(28,6){$K^-$}
      \put(27,27){$K^+$}
      \put(58,6){$K^-$}
      \put(69,27){$K^+$}
      \put(40,19){$\phi_1$}
      \put(64,19){$\phi_2$}
      \put(10,27){$\hat{n}_1$}
      \put(83,26){$\hat{n}_2$}
  \end{picture}
\end{center}
\label{fig:angles}
\caption{Helicity angles for the decay $B_s^0 \to \phi \phi$ where $\theta_{1,2}$ is the angle between the $K^+$ track momentum in the $\phi_{1,2}$ meson rest frame and the parent $\phi_{1,2}$ momentum in the $B^0_s$ rest frame, $\Phi$ is the angle between the two $\phi$ meson decay planes and $\hat{n}_{1,2}$ is the unit vector normal to the decay plane of the $\phi_{1,2}$ meson.}
\end{figure}
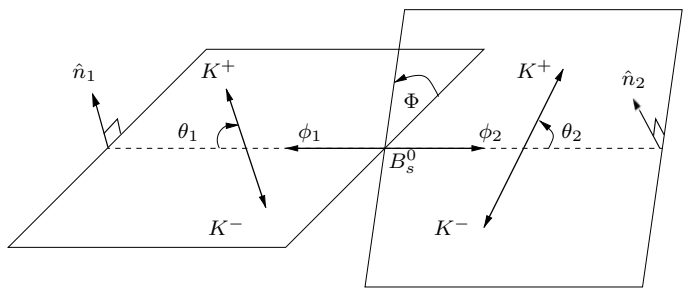

To measure polarisation amplitudes, a time-integrated untagged angular analysis is performed. Due to fast oscillation of the $B_s^0$ meson an equal number of $B_s^0$ and $\bar{B_s^0}$ are assumed at production. In addition, the $CP$ violating phase is assumed to be zero, as predicted in Ref.~\cite{raidal1}. Under these assumptions, the differential decay width is given
\begin{eqnarray}
\label{eqn:DiffDecRate}
\frac{32\pi}{9}\frac{d^3\Gamma}{d\cos\theta_1 d\cos\theta_2 d\Phi} = \frac{4}{\Gamma_L}|A_0|^2 \cos^2\theta_1 \cos^2\theta_2 + \frac{1}{\Gamma_L}|A_\parallel|^2 \sin^2 \theta_1 \sin^2 \theta_2 (1+\cos2\Phi) \nonumber \\
+ \frac{1}{\Gamma_H}|A_\perp|^2 \sin^2 \theta_1 \sin^2 \theta_2 (1-\cos2\Phi)
+ \frac{\sqrt{2}}{\Gamma_L}\textrm{Im}(A_0 A_\parallel^*)\sin2\theta_1\sin2\theta_2\cos\Phi.
\end{eqnarray}
where the helicity angles $(\theta_1, \theta_2, \Phi$) are defined in Fig.~\ref{fig:angles} and $\Gamma_{L,H}$ are the lifetimes for the light and heavy mass eigenstates respectively. The strong phase difference $\delta_\parallel$ is defined as arg($A_\parallel/A_0$).

Non zero triple product asymmetries can be due to either T violation or final state interactions. The former, assuming $CPT$ conservation, implies $CP$ is violated. In the decay {$B_s^0 \to \phi \phi$} two triple products can be constructed, denoted $U=\sin(2\Phi)/2$ and $V= \pm \sin(\Phi)$, where the positive sign is taken if the T-even quantity $\cos\theta_1\cos\theta_2 \geq 0$ and the negative sign otherwise. These correspond to the T-odd triple products
\begin{eqnarray}
\sin\Phi = (\hat{n_1}\times \hat{n_2})\cdot \hat{p_1}, \hspace{1.5mm}
\sin(2\Phi)/2 = (\hat{n_1}\cdot \hat{n_2})(\hat{n_1}\times \hat{n_2})\cdot \hat{p_1},
\end{eqnarray}
where $\hat{n_i}$ (i = 1,2) is a unit vector perpendicular to the $\phi$ decay plane and $\hat{p_1}$ is a unit
vector in the direction of the $\phi$ momentum in the $B_s^0$ rest frame (see Fig.~\ref{fig:angles}). 

Extraction of the triple products is a simple counting experiment which does not require tagging the flavour of the $B_s^0$ meson or time dependence. The asymmetries are defined as
\begin{eqnarray}
A_U = \frac{N_+ - N_-}{N_+ + N_-}, \hspace{1.5mm}
A_V = \frac{M_+ - M_-}{M_+ + M_-},
\end{eqnarray}
where $N_+$ ($N_-$) is the number of events with $U>0$ ($U<0$) and $M_+$ ($M_-$) is the number of events with $V>0$ ($V<0$). The dataset used for these analyses consists of $801 \pm 29$ {$B_s^0 \to \phi \phi$} candidates in 1~fb$^{-1}$ of data collected at the LHCb detector~\cite{Alves:2008zz} at a centre-of-mass energy $\sqrt{s} = 7$~TeV.
\begin{figure}[b]
\begin{center}
\includegraphics[scale=0.4]{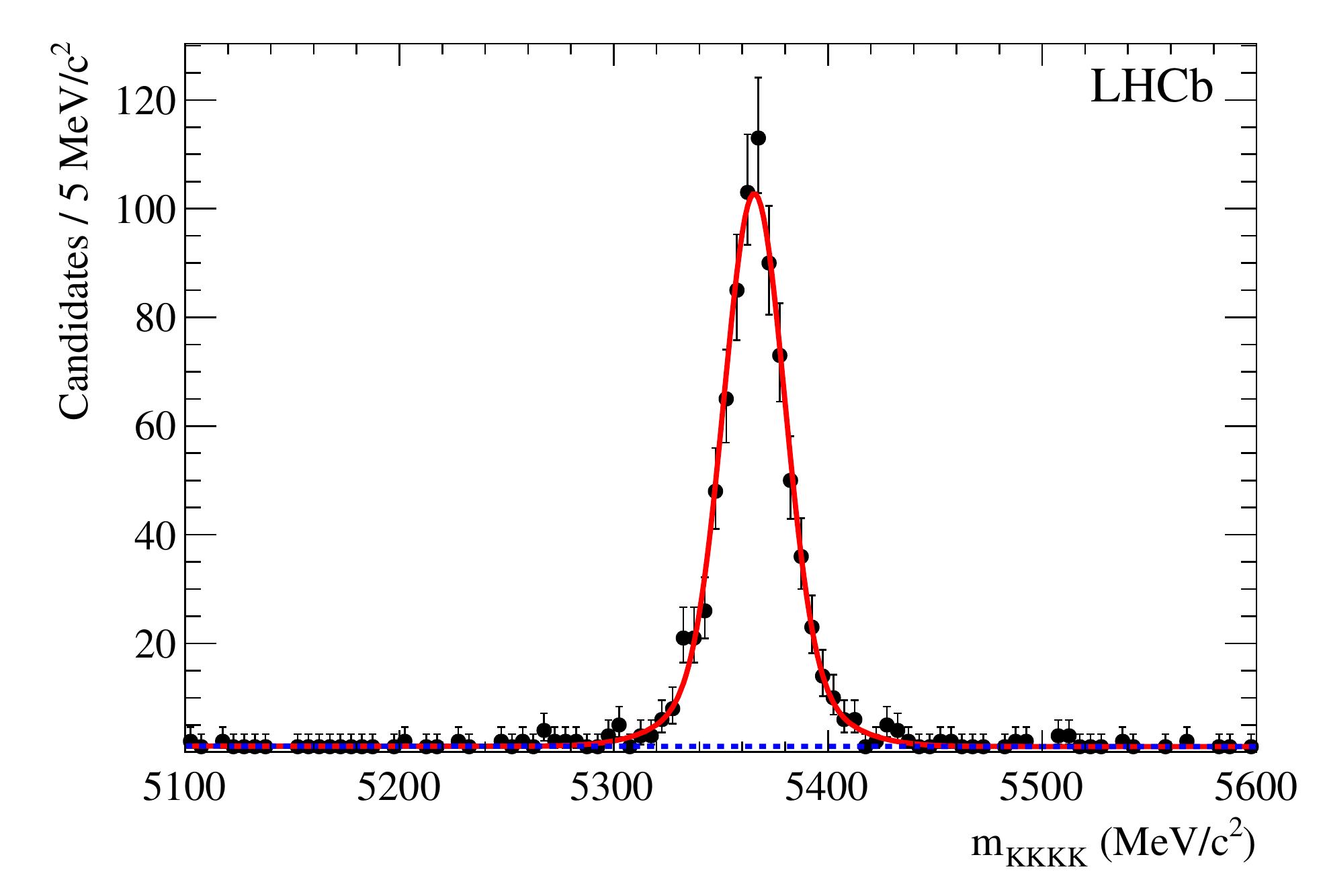}
\end{center}
\label{fig:mass}
\caption{The $K^+K^-K^+K^-$ invariant mass distribution for selected {$B_s^0 \to \phi \phi$} candidates.  A fit to
a double Gaussian signal component together with an exponential background (dotted line) is superimposed.}
\end{figure}
\section{Candidate Selection}
\label{sec:selection}
{$B_s^0 \to \phi \phi$} candidates are selected using events where both $\phi$ mesons decay into the final state $K^+K^-$. The candidate selection criteria were optimised using a data-driven approach based on the $_s\mathcal{P}lot$ technique~{\cite{splot}} with the invariant mass of the four-kaon system as the discriminating variable to separate signal from background. The figure of merit to be optimised is $S/\sqrt{S+B}$ where S (B) is signal (background) yield. 
%
Full details on the selection can be found in Ref.~\cite{phiphiTPA_CONF}.

Figure~\ref{fig:mass} shows the four-kaon mass for selected events. An unbinned maximum likelihood fit is used to extract the signal yield. The signal component is modelled by two Gaussian functions with a common mean. The relative fraction and width of the second Gaussian are fixed to values obtained from simulation. Combinatoric background is modelled by an exponential function. Background from $B^0 \to \phi K^{*0}(892)$ and $B_s^0 \to K^{*0}(892) \bar{K^{*0}}(892)$ is found to be negligible in both data-driven and simulation studies. Fitting this probability density function (PDF) yields $801 \pm 29$ signal events.

Data-driven studies to determine contributions from S-wave final states arising from $f_0 \to K^+K^-$ and non-resonant $K^+K^-$ yield results consistent with zero. The S-wave fraction is hence assumed to be zero in this analysis. A systematic error is assigned based on this assumption.
\section{Results}
\begin{figure}[ht]
\begin{center}
\subfigure{
\includegraphics[height=3.3cm]{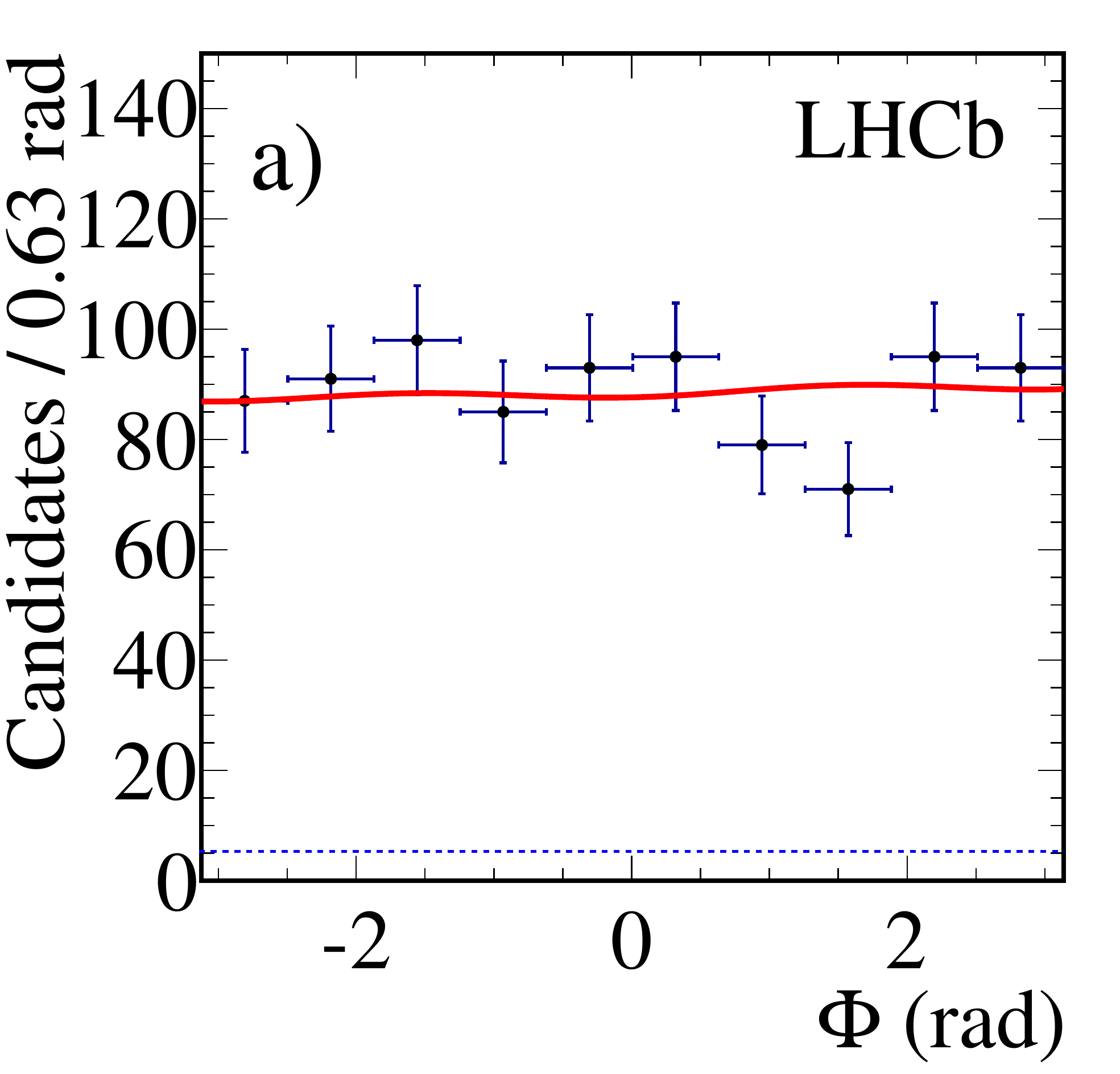}
}
\subfigure{
\includegraphics[height=3.3cm]{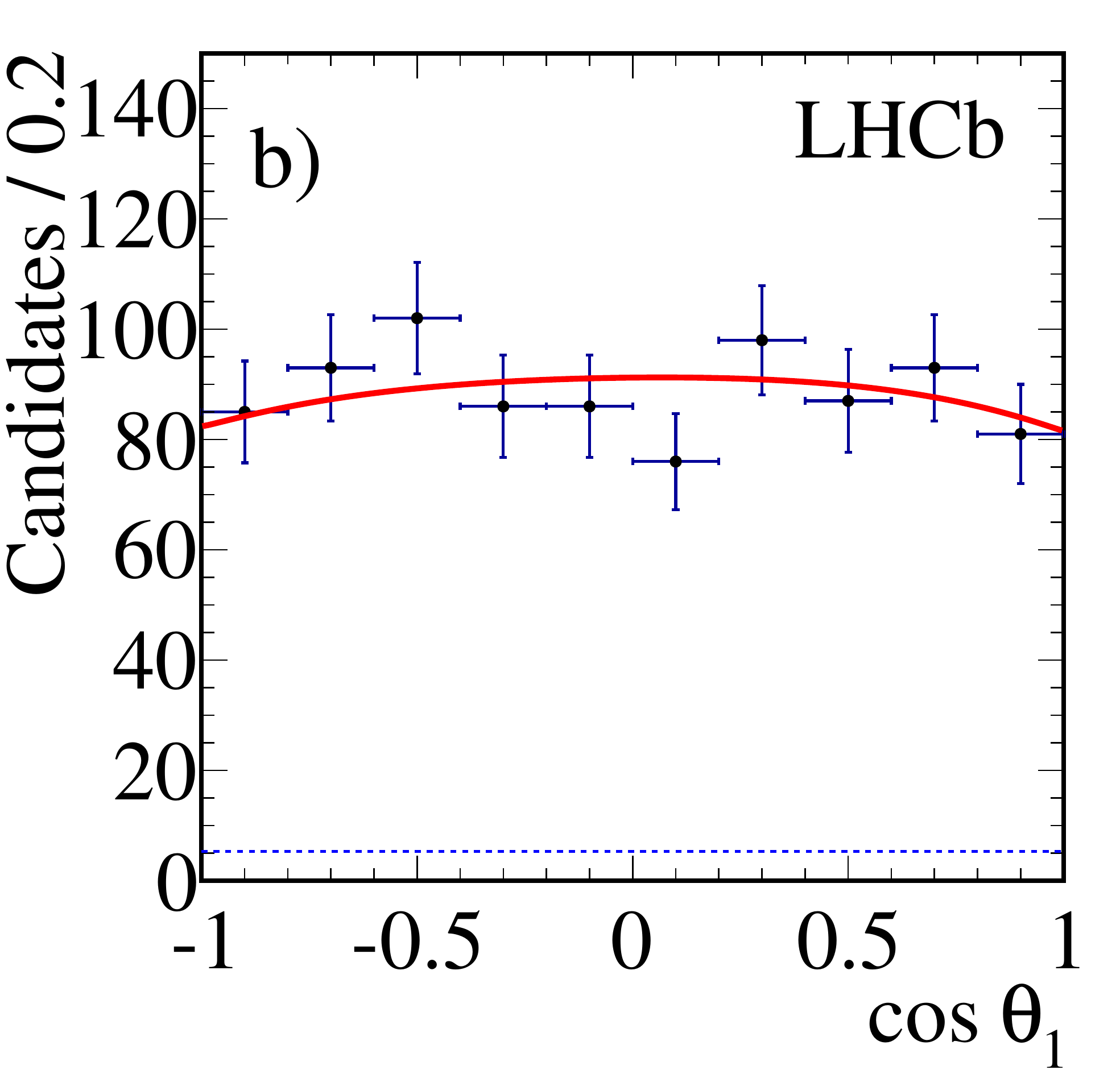}
}
\subfigure{
\includegraphics[height=3.3cm]{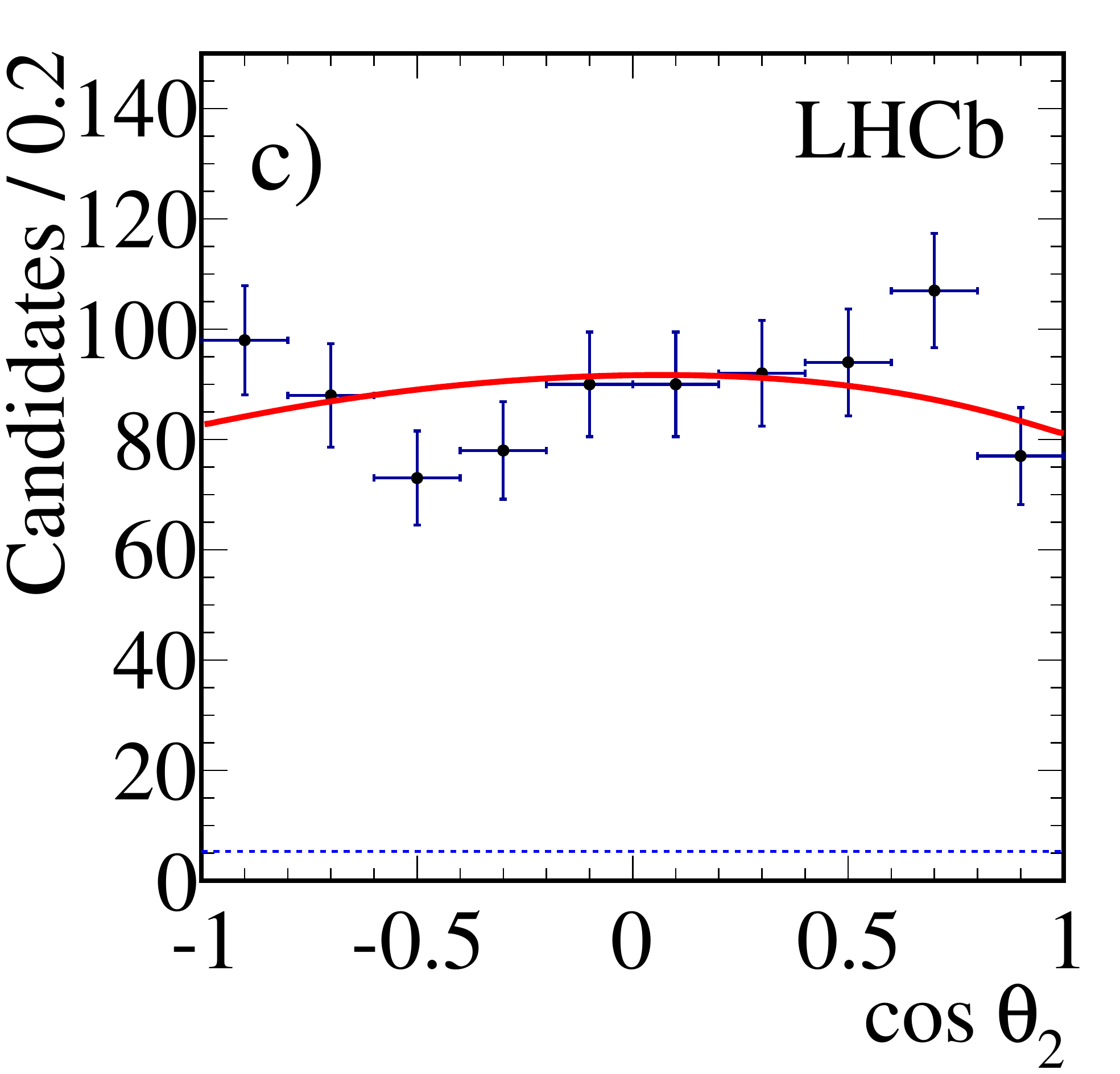}
}
\label{fig:fitAngles}
\end{center}
\caption{Angular distributions for (a) $\Phi$, (b) $\cos\theta_1$ and (c) $\cos \theta_2$ of {$B_s^0 \to \phi \phi$} events with the
fit projections for the total fitted PDF (solid line) and background component (dotted line) superimposed.}
\end{figure}
\begin{table}
\caption{Systematic uncertainties on the measured polarisation amplitudes and the strong
phase difference.\label{tab:syst_amp}}
\vspace{0.4cm}
\begin{center}
\begin{tabular}{|l|c|c|c|c|}
\hline
Source & $|A_0|^2$ & $|A_\perp|^2$ & $|A_\parallel|^2$ & $\cos\delta_\parallel$  \\
\hline 
{S-wave component} & {0.007} & {0.005} & {0.012} & {0.001} \\ 
Decay time acceptance  & 0.006 & 0.006 & 0.002 & 0.007 \\
{Angular acceptance} & {0.007} & {0.006}  & {0.006}  & {0.028} \\
Trigger category & 0.003 & 0.002 & 0.001 & 0.004 \\
Background model & 0.001 & - & 0.001 & 0.003 \\ \hline
Total & 0.012 & 0.010 & 0.014 &  0.029 \\ \hline
\end{tabular}
\end{center}
\end{table}
The polarisation amplitudes are determined by performing an unbinned maximum likelihood fit to the reconstructed mass and helicity angle distributions. Both the signal and background PDFs are the products of a mass component together with an angular component. The angular component of the signal is given by Eq.~\ref{eqn:DiffDecRate} multiplied by the angular acceptance of the detector, where the acceptance is determined using the simulation. The polarisation amplitudes are constrained such that $|A_0|^2 + |A_\parallel|^2 + |A_\perp|^2 = 1$. The angular distributions for the background have been studied using the mass sidebands in the data, these distributions are consistent with being flat in $(\cos\theta_1, \cos\theta_2, \Phi)$. A uniform angular PDF is therefore assumed for the background and more complicated shapes are considered as part of the systematic studies. The values of $\Gamma_s= 0.657 \pm 0.009 \pm 0.008$ ps$^{-1}$ and $\Delta\Gamma_s= 0.123 \pm 0.029 \pm 0.011$ ps$^{-1}$ together with their correlation coefficient of $-0.3$, as measured by LHCb,~\cite{JPsiPhi} are used as a Gaussian constraint. The angular projections are shown in Fig.~\ref{fig:fitAngles}.
\begin{table}[ht]
\caption{Systematic uncertainties on the triple product asymmetries, $A_U$ and $A_V$.\label{tab:syst_tpa}}
\vspace{0.4cm}
\begin{center}
\begin{tabular}{|l|c|}
\hline
Source & $A_U$ \& $A_V$ uncertainty \\ \hline
{Angular acceptance} & {0.009} \\
{Decay time acceptance} & {0.014} \\
Fit model & 0.005 \\ \hline
Total & $0.018$ \\ \hline
\end{tabular}
\end{center}
\end{table}

To determine the triple product asymmetries, the dataset is partitioned according to whether U (V) is less than or greater than zero. Simultaneous fits are performed to the mass distributions for each of the two partitions. In these fits, the mean and resolution of the Gaussian signal component together with the slope of the exponential background component are common parameters. The asymmetries are left as free parameters and are fitted for directly in the simultaneous fit.

The measurements of the polarisation amplitudes and triple product asymmetries are summarised in Table~{\ref{tab:results}}. Several sources of systematic uncertainty are considered, summarised in Tables~{{\ref{tab:syst_amp}} and {\ref{tab:syst_tpa}}. The measured values agree well with previous measurements by the CDF collaboration.~\cite{costa1} The triple product asymmetries are consistent with zero and hence no indication of T-odd asymmetries are observed with the present statistics.
\begin{table}[ht]
\caption{Measurements of the polarisation amplitudes and triple product asymmetries in the decay {$B_s^0 \to \phi \phi$}.\label{tab:results}}
\begin{center}
\vspace{0.4cm}
\begin{tabular}{|l|c|}
\hline
Parameter & Measurement \\ \hline
$|A_0|^2$                   & $0.365 \pm 0.022$(stat.)$ \pm 0.012$(syst.)  \\
$|A_\perp|^2$               & $0.291 \pm 0.024$(stat.)$ \pm 0.010$(syst.)  \\
$|A_\parallel|^2$           & $0.344 \pm 0.024$(stat.)$ \pm 0.014$(syst.)  \\
$\cos(\delta_\parallel)$    & $-0.844 \pm 0.068$(stat.)$ \pm 0.029$(syst.)  \\
$A_U$                       & $-0.055 \pm 0.036$(stat.)$ \pm 0.018$(syst.)  \\
$A_V$                       & $0.010 \pm 0.036$(stat.)$ \pm 0.018$(syst.) \\ \hline
\end{tabular}
\end{center}
\end{table}
\section*{References}
\bibliographystyle{h-physrev}
\bibliography{moriond}
\end{document}